\documentstyle[12pt]{article}
\begin{document}

\def\linadj#1{\normalbaselines
   \multiply\lineskip#1\divide\lineskip100
   \multiply\baselineskip#1\divide\baselineskip100
   \multiply\lineskiplimit#1\divide\lineskiplimit100}
\font\titlebs=cmbx10 scaled \magstep3
\font\nama=cmbx10 scaled\magstephalf
\font\gbr=cmbx12

\linadj{200}

\noindent {\titlebs A unified model for temperature dependent electrical conduction in polymer electrolytes\\}

\noindent {\nama Mikrajuddin, I. Wuled Lenggoro, and Kikuo Okuyama$^{a}$}

\noindent {\sl Department of Chemical Engineering, Hiroshima University, Kagamiyama, Higashi-Hiroshima 739-8527, Japan.\\}

\begin{center}
{\bf Abstract}
\end{center}

The observed temperature dependence of electrical conduction in polymer electrolytes is usually fitted with two separated equations: an Arrhenius equation at low temperatures and Vogel-Tamman-Fulcher (VTF) at high temperatures. We report here a derivation of a single equation to explain the variation of electrical conduction in polymer electrolytes at all temperature ranges. Our single equation is in agreement with the experimental data.

\clearpage

There is a wide interest in composites of ceramic nanoparticles and solid polymer electrolytes (SPE) for their potential applications in batteries, fuel cells, solar cells, and other electrochemical devices [1-4]. Usually, the SPE only exhibits a high electrical conduction at high temperatues, i.e., above the melting point. However, dispersing ceramic nanoparticles in the SPE matrix can enhance the conductivity even at low temperatures by three order of magnitudes. The role of the dispersed nanoparticles is to influence the recrystalization kinetics of the SPE polymer chain, thereby ultimately promoting localized amorphous regions and thus enhancement of the ion transport [2]. A number of researchs have been performed to investigate the enhancement of conductivity in nanoparticles/SPE composites at room temperature to a practically useful values (of about $10^{-4}$ S/cm) [2-7].

Understanding the temperature dependent conductvity of SPE is strictly important in order to realize the practically useful nanoparticles/SPE composites. Indeed, it was demonstrated that the general behavior of conduction in nanoparticles/SPE composites is similar to that in pure SPE [2-4,7]. However, a detailed understanding of the conductive mechanism is still lacking [8,9].

The conductivity of SPE is attributed by cations transfer. At low temperatures, where $kT$ is far below the activation energy, the concept of ion conduction being the result of infrequent individual ion hops over large energy barriers is an appropriate picture of the conduction dynamics. This picture exhibits an Arrhenius temperature dependent conductivity, i.e., $\sigma_1(T) = (A_1/T)\,\exp [-E_a/kT]$. At higher temperatures, however, the effect of temperature would no longer be thermally creating charge carriers for conduction, but rather to increase the mobility of charge carriers. This hypotesis has been used to explain the nonlinearity of conductivity in glass materials [9]. The ionic motion is strongly coupled to the polymer segmental relaxation to enhance the mobility. Still there is no single theory that is able to explain the temperature dependent electrical conduction in SPE for all temperatures. In the present work we develop a model that describes the electrical conduction properties in polymer electrolytes at all temperature ranges.

In Fig. 1, the SPE at arbitrary temperature is divided into a number of similar cells. Each cell contains a number of clusters (chain segment) and each cluster is governed by a number of sites. Consider a certain cluster containing $m$ sites: a part is "amorphous" sites and the rest are "crystalline" sites. Other authors classify as "solid" atoms and "liquid" atoms [10]. The average interaction energy per site can be written as $E_m = - (J_o/z m)\, \sum_{i,j}\,\sigma_i \sigma_j$, with $\sigma_i = \pm 1$ denotes two states: "amorphous" and  "crystalline", $z$ is the number of nearest neighbors sites, and $J_o$ as a constant (interaction strength) [11].

Suppose there are $\ell$ "amorphous" and $m - \ell$ "crystalline" sites in a cluster of $m$ sites. The interaction energy per site (Bragg-Williams formalism) is $e_m(\ell) = e(L) = - J_o L^2/2$, with $L = 2 \ell/m - 1$ and the partition function is $Z_m = \sum_{\ell = 0}^m\,\{m!/(m-\ell)! \ell!\}\,\exp [-m e_m(\ell)/kT]$. Using Stirling approximation and converting the summation into integral, one finds $Z_m \cong \int_{-1}^{+1}\, dL\, \exp [-m f(L)/kT]$, with $f(L) = e(L) +(kT/2)\, \{ (1+L) \ln [(1+L)/2] + (1-L) \ln [(1-L)/2] \}$. If $m$ is sufficiently large, only the minimum free energy contributes significantly to the integration. This appears at $L = L_\infty$ that satisfies $\partial f/\partial L = 0$ or $L_\infty = \tanh \left [{L_\infty J_o/kT} \right ]$. For $T > J_o/k$, the only solution is $L_\infty = 0$. This standard result is only valid for the macroscopic limit [11]. For a mesoscopic system, the energy can be lowered as $\bar{e}_m \approx \,\int_{-1}^{+1}\,dL(-J_o L^2/2)$ $\exp[-m f(L)/kT]/Z_m$. For large but finite $m$ the integral may be approximated by a steepest descent procedure and found that $\bar{e}_m \approx -J_o/2(1 - Jo/kT)$. The energies of clusters are similar regarless of their size. The energy of a cell containing $n$ clusters can be approximated as $E_n = - nJ_o/2(1 - Jo/kT)$.

The instant velocity of ion in a SPE is $\vec{v}_i = \vec{v}_{ip} + \vec{v}_p$ with $\vec{v}_{ip}$ the velocity of ion relative to polymer chain and $\vec{v}_p$ the velocity of polymer chain. The average kinetic energy of ion become $\langle KE \rangle$ = $m_i \langle \vec{v}_{ip}^2 \rangle/2$ + $m_i \langle \vec{v}_{ip}.\vec{v}_p \rangle$ + $m_i \langle \vec{v}_{p}^2 \rangle/2$. Since the ions are strongly coupled to polymer chain, the ions follow the fluctuation of polymer chain. And since the segmental motion of chain dominates at high temperatures, one has $\vert \vec{v}_{ip} \vert \ll \vert \vec{v}_p \vert$, so that $\langle KE \rangle \propto \langle \vec{v}_{p}^2 \rangle$, i.e., proportional to the kinetic energy of polymer in a cell, $E_n$. The total energy of ion in a cell containing $n$ cluster can be approximated by $E_n^i = V_i + \kappa E_n$, with $\kappa$ ia a real constant. The potential energy of ion, $V_i$, is dominated by interaction of cation and anion and the interaction with the nearest atom in polymer chain at which the cation is attached. The potential energy therefore can be considered to be constant, independent of temperature and chain fluctuation.

To describe the effect of temperature on conductivity, let us start by investigating the conductance between cells. A mixture of amorphous and crystalline cells is similar to the mixture of conducting filler and insulator matrix. The conducting filler is associated with the amorphous cell and the matrix is associated with the crystalline cells. The increase in temperature resulting the increase in the fraction of amorphous state is similar to the increase in the volume fraction of conduction filler loaded in insulator matrix.

Firstly we calculate the fraction of amorphous state. Due to fluctuation, there is a probability for the melting process to occur at an arbitrary temperature. Assuming a Gaussian type of fluctuation, the probability for melting in the range of temperature between $T$ and $T+dT$ is $w(T) dT$ = $[1/\sqrt{2 \pi \langle (\Delta T)^2 \rangle_m}]$ $\exp \left [{-(\Delta T)^2/2 \langle (\Delta T)^2 \rangle_m} \right ] d \Delta T$, where $T_m$ is the commonly observed melting point, $\Delta T = T - T_m$, $d \Delta T$ = $d(T - T_m)$ = $dT$ and $\langle (\Delta T)^2 \rangle_m$ is the assembly average of $(\Delta T)^2$ at $T_m$ [12]. The probability for the presence of the amorphous state at temperature $T$ is equal to the probability for attaining melting points at all temperatures below $T$, i.e. $p(T)$ = $\int_{- \infty}^T\,[1/\sqrt{2 \pi \langle (\Delta T)^2 \rangle_m}]$ $\exp \left [{-(\Delta T)^2/2 \langle (\Delta T')^2 \rangle_m} \right ] d \Delta T'$ or
\begin{equation}
p(T) =  \cases {1/2 -1/2\, {\rm erf} \left [{- \Delta T/\sqrt{\langle (\Delta T)^2 \rangle_m}} \right ] & if $T < T_m$,\cr
               \cr
               1/2 +1/2 {\rm erf} \left [{\Delta T/\sqrt{\langle (\Delta T)^2 \rangle_m}} \right ] & if $T > T_m$.\cr}
\end{equation}
with $ {\rm erf}(x)$ = $(2/\sqrt{\pi})$ $\int_0^x$ $\exp[-t^2/2] dt$. Since $\langle (\Delta T)^2 \rangle = R T^2/C_v$ with $C_v$ is the molar heat capacity at constant temperature [12], one can approximate $\langle (\Delta T)^2 \rangle_m \cong R T_m^2/C_{vm*}$ with $C_{vm*}$ is the average molar heat capacity at $T_m$ and
\begin{equation}
p(T) = \cases {1/2 - 1/2\, {\rm erf}[-C_{vm*} \Delta T/R] & if $T < T_m$,\cr
               \cr
               1/2 + 1/2\, {\rm erf}[ C_{vm*} \Delta T/R] & if $T > T_m$.\cr}
\end{equation}
This probability denotes the fraction of amorphous phase, i.e., $w_a(T) = p(T)$ [13].

Based on Fig. 1, define $f$ as the coordination number (cell hands) and $\alpha$ the probability for direct bonding of two cells. In a polycell containing $n$ cells ($n$-cell), some hands formed bonds and the others remain free. It requires the attachment of $(n-1)$ new cells following a preselected cell to form $n$-cell. If only one bond is created at each attachment of a new cell, $(n-1)$ bonds and $(f-2)n + 1$ free hands (excluded a freely preselected hand) appeared having the $n$-cell been formed. Since the probability of hand to form bond is $\alpha$ and to become free is $(1 - \alpha)$, the total probability for creating $n$-cell of any configuration is
\begin{equation}
P_n = \Omega_n \alpha^{n-1} ( 1 - \alpha)^{(f - 2) n + 1}
\end{equation}
with $\Omega_n$ is the total number of configurations.

To form $(n-1)$ bonds in an $n$-cell, it allows to select $(n-1)$ from the total ($n-1)f$ hands belong to the added cells in $\left [{f(n - 1)} \right ]!/{\left [{(f -2) n + 1} \right ]! (n - 1)!}$ ways, and the added cells can be attached sequentially in $(n - 1)!$ ways. Noting that all cells in the $n$-cell are identical, so that we have to add a division factor $n!$ and obtaining the expression $\Omega_n$ = $\left [{f(n - 1)} \right ]!/{\left [{(f -2) n + 1} \right ]!(n-1)!}$ $\times (n - 1)!/n!$ = $\left [{f(n - 1)} \right ]!/\left [{(f -2) n + 1} \right ]! n!$.

The alternative expression for $P_n$ is $P_n$ = {\sl Number of free hands on} $n$-{\sl cell/The total number of free hands} [14]. The number of free hands belongs to an $n$-cell (the preselected cell plus the added cells) is $(f - 2)n + 2$. Suppose $N_n$ denotes the population of $n$-cell and $N_o$ the number of all cells. The total number of hands belongs to all cells is $N_o f$ and the total number of free hands is $N_o f (1 - \alpha)$. Therefore
\begin{equation}
P_n = \left [{(f - 2) n + 2} \right ] N_n/{N_o f (1 - \alpha)}
\end{equation}
From Eqs (3) and (4) one obtains $N_n = N_o f[(1 - \alpha)^2 / \alpha] \Omega_n \beta^n$, with $\beta = \alpha (1 - \alpha)^{f - 2}$.

The weight fraction of $n$-cell is $w_n  = n N_n/N_o$ = $f[(1 - \alpha)^2/\alpha] n \Omega_n \beta^n$, with $\sum_{n = 1}^{\infty}\,w_n$ = $\sum_{n = 1}^{\infty}\,f[(1 - \alpha)^2/\alpha] n \Omega_n \beta^n = 1$. The summation result on the right hand side depends on $\beta$. For a specific value of $\beta$, there is only one summation result. Since $\beta$ is a polynomial function of $\alpha$, a specific value of $\beta$ results more than one value of $\alpha$. By replacing $\alpha$ with $\alpha'$ that also satisfies $\beta$ = $\alpha' (1 - \alpha')^{f-2}$, the summation result on the right hand side never changes. However, since in general $(1 - \alpha)^2/\alpha$ $\ne$ $(1 - \alpha')^2/\alpha'$, we have, in general, $\sum_{n = 1}^{\infty}\,w_n \ne 1$, that sounds unphysical. The physical soundness is achieved only when the smallest root of $\beta = \alpha (1 - \alpha)^{f - 2}$ is used [14]. Therefore, the accepted expression for $w_n$ would be $w_n =  f[(1 - \alpha')^2/\alpha'] n \Omega_n \beta^n$ with $\alpha'$ is the smallest root of $\alpha' (1 - \alpha')^{f - 2}$ = $\alpha (1 - \alpha)^{f - 2}$.

The summation of $\Omega_n$ for all finite $n$ gives the weight fraction of the so called sol with respect to the amorphous phase, $w_s = \sum_{\rm all\,\, finite\,\, n}w_n$ = $(1 - \alpha)^2 \alpha'/(1 - \alpha')^2 \alpha]$. The weight fraction of the so called gel (infinity network) with respect to the amorphous phase is then
\begin{equation}
w_g = 1 - (1 - \alpha)^2 \alpha'/(1 - \alpha')^2 \alpha.
\end{equation}
The weight fraction of gel with respect to the total weight of SPE is 
\begin{equation}
w_g' = w_a w_g.
\end{equation}

Since the presence of segmental fluctuation, the number of cluster is random in all cells to result the randomness in ionic energy of cells. The amorphous cells also occupy random positions in a SPE. Both energy and position of amorphous cells satisfy the requirement for the application of critical path method approach[15] so that the conductance between two cells having $n$ and $n'$ clusters can be written as $g_{n,n'} = (g_o/T)\, \exp [-2 \bar{\gamma} s_{nn'} - W^i_{nn'}/kT]$, with $g_o$ is the prefactor, $\bar{\gamma}$ is the screening parameter, $s_{nn'}$ is the distance between two cells, and $W^i_{nn'} = \kappa \,(\vert n \bar{e}_n \vert + \vert n' \bar{e}_{n'} \vert + \vert n \bar{e}_n - n' \bar{e}_{n'} \vert)/2$.

We only need to consider the connected amorphous cells since only them contribute to the electrical conduction. Since the amorphous cells are connected at a constant distance (for nearest neighbor ones), the first term in exponential is constant. Sheng, Sinchel, and Gittleman argued that the temperature dependence of conductivity of composite consists of carbon particles dispersed in polyvinylchloride is similar to that of between two carbon particles in that composites [16]. Following this argument, the conductivity of PE (at high temperatures) is given by
\begin{equation}
\sigma_2(T) \cong {A_2 \over T}\,\exp [-B/k (T - T_o)]
\end{equation}
with $B$ is taken to be the average of $(\vert n - n' \vert + n + n')$ multiplied by $\kappa J_o$, and $T_o = J_o/k$, which proves that at high temperatures, the conductivity varies according to VTF law [8,17].

Although the conductivity of sol is higher than that of crystalline phase, since it is always covered by crystalline phase, the effect of sol on conductivity of the PE is ignorable. It is analog to insulator coated conducting particles, where the presence of insulator on the surfaces blocks the transport of charges. Practically, the polymer can be classified into two phases: a gel phase with conductivity $\sigma_2(T)$ and a mixture of crystalline phase and sol with a conductivity of $\sigma_1(T)$. The total conductivity of polymer is then
\begin{equation}
\sigma(T) = (1 - v_g) \sigma_1(T) + v_g \sigma_2(T)
\end{equation}

As an illustration let us use the above result to explain the experimental observation of the electrical conductivities of polyethylene oxide (PEO) based polymer electrolytes. We select cubical shapes of cell such that it contact with six other cell ($f = 6$). The probability for direct bonding of two cells $\alpha$ is equal to $w_a$, the volume fraction of amorphous phase. The heat capacity of polyethylene oxide at constant pressure is around $C_p \cong 30$ J/mol K [18]. For relatively narrow temperature range one can assume that heat capacity only changes very little with temperature (can be considered to be constant). Using a rough approximation (that usually used for gas at low pressure), i.e., $C_v \cong C_p - R$, we used in the calculation $C_{vm*} \cong 25$ J/mol K.

Figure 2 shows the comparison of the present model and experimental data for a system of: (a) Cd$_{0.75}$PS$_3$Li$_{0.5}$(PEO) [8], (b) Cd$_{0.75}$PS$_3$Na$_{0.5}$(PEO) [8], (c) Cd$_{0.75}$PS$_3$Li$_{0.5}$(PPG) [17], with PPG denotes polypropylene glycol. The parameters used in calculation are displayed in Table 1. The values of those parameters are close to that used by Jeevanandam and Vasudevan to fit the experimental data using two separated curves: an only-Arrhenius curve for low temperatures conductivity and only-VTF curve at high temperatures [8,17]. Indeed, these values often depend on the preparation condition. Eventually, similar material provides different values of these parameters. The present model results in a single equation for the conductivity behavior of solid polymer electrolytes and also successfully explains the change in electrical conductivity from Arrhenius like at low temperatures to VTF like at high temperatures with a continuous slope.

We will show that the parameters used in calculations are acceptable. By nothing that $w_a = \alpha$, it can be shown that if $w_a < 1/(f - 1)$, $w_g = 0$. According to Eq. (8), this condition implies that only Arrhenius behavior to occur. When $w_a > 1/(f - 1)$, $w_g > 0$ so that the VTF behavior occurs. Thus the temperature at which  $w_g$ changes from zero to non-zero can be considered as the critical temperature, at which the VTF behavior of conductivity starts to occur and becomes dominant by further increasing the temperature. This temperature is known as the glass temperature, $T_g$, that satisfies $w_a(T_g) = 1/(f-1)$. Since the glass temperature is less than the melting point then using Eq. (2) one has $1/(f - 1)$ = $1/2 - 1/2\,{\rm erf}[\sqrt{C_{vm*}/R}(T_m - T_g)/T_m]$, or
\begin{equation}
T_g = T_m \left ({1 - \sqrt{C_{vm*} \over R}\,{\rm erf} \left [{1 - {2 \over f - 1}} \right ]} \right )
\end{equation}
Using $f = 6$ and $C_{vm*} = 25$ J/mol K, we have $T_g/Tm$ = $0.68$ The ratio of $T_g/T_m$ observed on a lot of polymers including polyethylene, polypropylene, polystyrene, PEO are located between $0.5$ and $0.8$ [19]. Since our result is located in the range of observation date, the selected parameters $C_{vm*}$ as well as $f$ are acceptable.

To verify the acceptability of parameters $T_m$ and $T_o$, let us use the relation between $T_g$ and $T_o$ (Vogell temperature) as $T_g$ = $T_o$ + $50$ K [17]. Using $T_o$ in Tabel 1, the glass temperature of the three types polymer electrolytes are $140.7$ K, $140.7$ K, and $192.2$ K for Cd$_{0.75}$PS$_3$Li$_{0.5}$(PEO), Cd$_{0.75}$PS$_3$Na$_{0.5}$(PEO), and Cd$_{0.75}$PS$_3$Li$_{0.5}$(PPG), respectively. Furthermore, using $T_m$ in the last column of Tabel 1, we have the ratio of $T_g$ and $T_m$ for these three SPE are $0.6544$, $0.6544$, and $0.6537$ for Cd$_{0.75}$PS$_3$Li$_{0.5}$(PEO), Cd$_{0.75}$PS$_3$Na$_{0.5}$(PEO), and Cd$_{0.75}$PS$_3$Li$_{0.5}$(PPG), respectively. Again, these values are also located in the experimental observation, proved that the selected $T_o$ and $T_m$ are acceptable.

It is obvious from Eq. (9) that the glass temperature is located near the melting point if $C_v$  is very large. This behavior can be understood easily from eq. (2) and the relation  $\langle (\Delta T)^2 \rangle = R T^2/C_v$, where  $C_v$ is inversely proportional to the standard deviation of fluctuation, thereby the large value of $C_v$  implies to the small value of standard deviation such that the melting point is localized in a small range of temperatures. Since the factor   generally depends on the type of polymer, the ratio of $T_g/T_m$ is also polymer specific.

In conclusion, we have shown that by considering the fluctuation of melting point and the application of gelation theory the determine the formation of amorphous states network we can explain the presence of two conductivity behavior in polymer electrolytes. We showed that the VTF behavior at high temperatures can be derived using a simple Ising model. Our result in the form of a single equation for electrical conductivity at all temperature ranges is in agreement with several experimental evidences. Our results are easily extended to explain the conductivity behavior of nanoparticles/SPE composites.

A scholarship provided by the Japanese Ministry of Education, Science, Sports, and Culture for Mikrajuddin is gratefully acknowledged.\\

\noindent E-mail address: okuyama@hiroshima-u.ac.jp

\noindent [1] P.G. Bruce (Ed.), Solid State Electrochemistry, Cambridge University Press, London, 1995.

\noindent [2] F. Croce, G. B. Appetecchi, L. Persi, and B. Scrosati, Nature {bf 394}, 456 (1998).

\noindent [3] F. Capuano, F. Corce, and B. Scrosati, J. Electrochem. Soc. {\bf 138}, 1918 (1991).

\noindent [4] W. Wieczorek, Mater. Sci. Eng. {\bf B15}, 108 (1992).

\noindent [5] M. C. Borghini, M. Mastragostino, S. Passerini, and B. Scrosati, J. Electrochem. Soc. {\bf 42}, 2118 (1995).

\noindent [6] B. Kumar and L. Scanlon, J. Power Source {\bf 52}, 261 (1994).

\noindent [7] W. Wiczorek, Z. Florajanczyk, and J. R. Stevens, Electrochim. Acta {\bf 40}, 2251 (1995).

\noindent [8] P. Jeevanandam and S. Vasudevan, J. Chem. Phys. {\bf 109}, 8109 (1998)

\noindent [9] J. Kincs and S. W. Martin, Phys. Rev. Lett {\bf 76}, 70 (1996).

\noindent [10] V. N. Novikov, E. Rosler, V. K. Malinovsky, and N. V. Surotsev, Europhys. Lett. {\bf 35}, 189 (1996).

\noindent [11] R. V. Chamberlin, Phys. Rev. Lett. {\bf 82}, 2520 (1999).

\noindent [12] L. D. Landau, E. M. Lifshitz, Statistical Physics, 3rd ed., Butterworth-Heinemann, Oxford, 1997.

\noindent [13] Mikrajuddin, F.G. Shi and K. Okuyama, Microelectron. J. {\bf 31} 261 (2000).

\noindent [14] P. J. Flory, Principles of Polymeric Chemistry (Cornel University Press, New York 1956).

\noindent [15] V. Ambegaokar, B. I. Halperin, and J. S. Langer, Phys. Rev. {\bf B4}, 2612 (1971).

\noindent [16] P. Sheng. E. K. Sinchel, and J. I. Gittleman, Phys. Rev. Lett. {\bf 40}, 1197 (1978).

\noindent [17] P. Jeevanandam and S. Vasudevan, J. Phys. Chem. B {\bf 102}, 4753 (1998)

\noindent [18] J. Branrup and E. H. Emergut (Eds), Polymer Handbook 3 (John Wiley, New York, 1989).

\noindent [19] R. J. Young and P. A. Lovell, Introduction to Polymers 2nd ed. (Chapman \& Hall, London 1991), p. 297.

\clearpage

\begin{center}
{\bf Figure Captions\\}
\end{center}

\noindent Figure 1.

The modeling steps used in this work: (a) At arbitrary temperature the SPE always contains amorphous and crystalline phase. (b) The SPE is divided into a number of similar cells. The content of each cell is either dominated by amorphous phase or crystalline phase. We can consider the SPE as a "solution" of amorphous cells into a "solvent" of crystalline phase. Some amorphous cells form network. Networks that contain finite number of amorphous cell is named as "sol" and that contain infinity number of amorphous cells is named as gel. (c) The formation of cells network can be similarized as the polymerization process. The cell denoted by (1) is the preselected one and the marked cell hand denotes the preselected hand.\\

\noindent Figure 2.

The comparison of the theory calculation (lines) and the experimental data (symbols) for systems of (a) Cd$_{0.75}$PS$_3$Na$_{0.5}$(PEO) [8], [open circle, solid line], (b) Cd$_{0.75}$PS$_3$Li$_{0.5}$(PEO) [8], [plus symbol, dotted line] and (c) Cd$_{0.75}$PS$_3$Li$_{0.5}$(PPG) [17] [cross, dashed line].\\\\

\clearpage

\noindent Table 1. The parameters used to calculate the theoretical lines in Figure 2.
$$ \vbox{\offinterlineskip\vskip3pt
   \smallskip\hrule\smallskip
   \halign {#\hfil &\quad# &&\quad\hfil#\hfill\cr
   Figure &\vrule& $A_1$ (S K/cm) & $E_a$ (eV) & $A_2$ (S K/cm)& $B$ (eV)& $T_o$ (K) & $T_m$ (K)\cr
   \noalign{\smallskip\hrule\smallskip}
 2(a) &\vrule & $1.07 \times 10^{-8}$ & $0.052$ & $35$ & $0.195$ & $90.7$ & $215$\cr
 2(b) &\vrule & $2.32 \times 10^{-5}$ & $0.130$ & $125$ & $0.265$ & $90.7$ & $215$\cr
 2(c) &\vrule & $1.05 \times 10^{-2}$ & $0.197$ & $0.97$ & $0.125$ & $142.2$ & $294$\cr
  \noalign{\smallskip\hrule\smallskip} }}$$

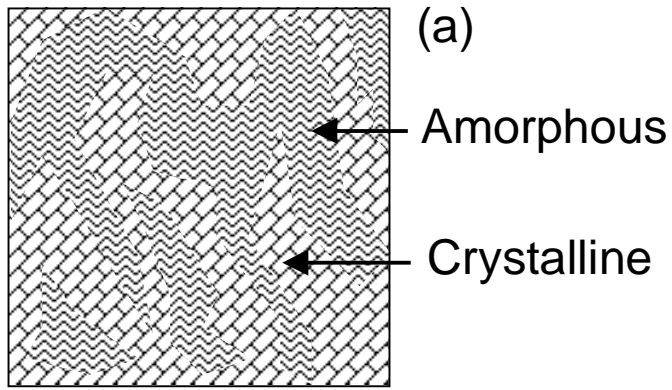
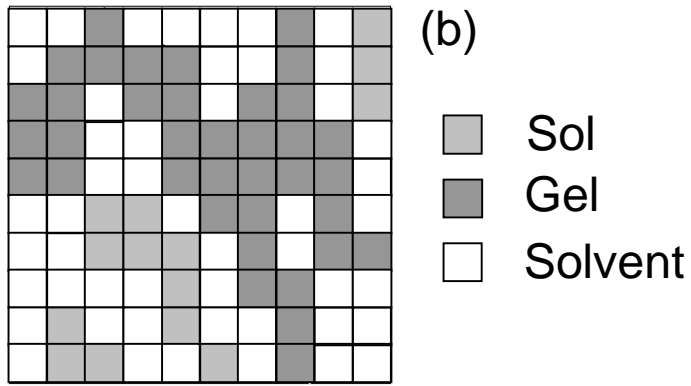
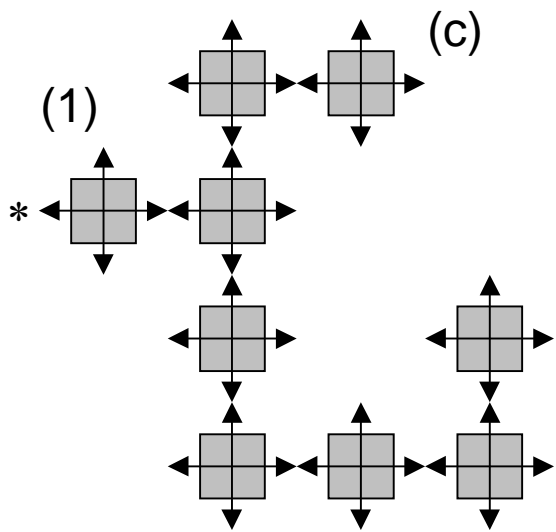
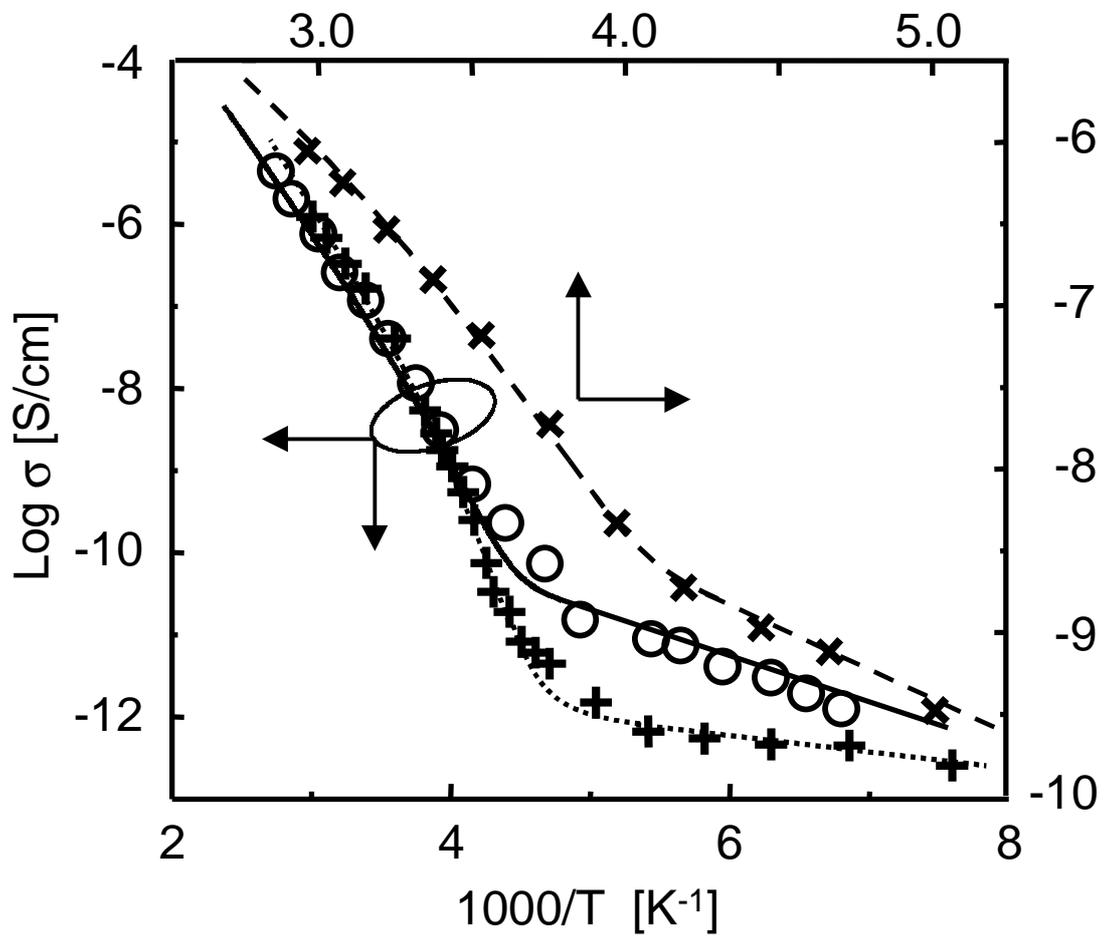
\end{document}